\documentclass[%
 reprint,
superscriptaddress,
 amsmath,amssymb,
 aps,
 pre
]{revtex4-2}
\usepackage[dvipdfmx]{graphicx}
\usepackage[dvipdfmx]{xcolor}
\usepackage{dcolumn}
\usepackage{bm}

\begin{document}

\title{
Emergence of compact-disordered phase in a polymer Potts model
}
\author{Ryo Nakanishi}
\affiliation{
 Graduate School of Arts and Sciences,
 The University of Tokyo,
 Komaba, Meguro-ku, Tokyo 153-8902, Japan
}

\author{Koji Hukushima}
\affiliation{
 Graduate School of Arts and Sciences,
 The University of Tokyo,
 Komaba, Meguro-ku, Tokyo 153-8902, Japan
}
\date{\today}

\begin{abstract}
One of the central problems in epigenetics is how epigenetic modification patterns and chromatin structure are regulated in the cell nucleus. The polymer Potts model, a recently studied model of chromatins, is introduced with an offset in the interaction energy as a parameter, and the equilibrium properties are investigated using the mean-field analysis of the lattice model and molecular dynamics simulations of the off-lattice model. The results show that in common with both models, a phase emerges, which could be called the compact-disordered phase, in which the polymer conformation is compact and the epigenetic modification pattern is disordered, depending on the offset in the interaction energy and the fraction of the modified nucleosomes.
\end{abstract}

\maketitle

\section{Introduction}
 Epigenetics is defined as ``the study of mitotically and/or meiotically heritable changes in gene function that cannot be explained by changes in DNA sequence"\cite{felsenfeldBriefHistoryEpigenetics2014}, and the fundamental problem in the field is to understand how a single fertilized zygote develops into a mature organism. In eukaryotes, genomic DNA is wrapped around histone octamers to form nucleosomes, which are strung together to form chromatin\cite{maeshimaPhysicalNatureChromatin2021} and it has become evident that the chromatin structure and dynamics, as well as biochemical modifications of DNA and histones, play important roles in epigenetic regulation\cite{felsenfeldBriefHistoryEpigenetics2014}. 
 Chromatin can be classified into two regions. The one in which genes are actively transcribed is called euchromatin and the other in which genes are repressed is called heterochromatin\cite{HENIKOFF2000O1,poonpermFormationMultiLayered2021}. Depending on the chromatin regions, histone proteins have distinct epigenetic modifications. 
 
 Several studies have attempted to understand various phenomena related to epigenetics through mathematical modeling approaches. From a physics perspective, many theoretical works have focused on the multistability of the epigenetic marks, their spatial patterns, and their heritability\cite{RevModPhys.88.025002}. The establishment of epigenetic modification patterns was initially studied using one-dimensional mathematical models\cite{DODD2011624, anink-groenenMechanisticStochasticModel2014a, RevModPhys.88.025002}. More recently, several polymer models have been developed to study the coupling between the one-dimensional epigenetic modification pattern along the chromatin and three-dimensional polymer dynamics\cite{PhysRevX.6.041047, sandholtzPhysicalModelingHeritability2020, KATAVA20222895, ABDULLA2023102033}. For example, the ``polymer Potts model" or ``magnetic polymer"\cite{PhysRevX.6.041047, PhysRevE.100.052410, PhysRevLett.123.228101} has been proposed for a chromatin model. It explicitly includes the microscopic degrees of freedom of polymer conformation and nucleosome modifications simultaneously.
Such mathematical models have been studied from the perspective of macroscopic phase transition phenomena and thermodynamics using statistical mechanics tools such as molecular dynamics simulations.

Magnetic polymer models, consisting of monomers with magnetic moments, have been studied with attention to the magnetic properties of polymer materials\cite{PhysRevB.13.2186}. Moreover, these have recently been studied actively as abstract models of chromatin\cite{PhysRevX.6.041047,PhysRevE.100.052410,PhysRevLett.123.228101,AdachiKawaguchi2019}. In this context, a single polymer model with a chain of $N$ monomers representing nucleosomes, the structural units of chromatin, is often used. Corresponding to the fact that the nucleosome has various modification states, each monomer has a ``spin" as an internal degree of freedom. With $i$ as the index of the monomers in the chain, the position and spin of the $i$-th monomer are represented by $\vec r_i$, a vector in three spatial dimensions, and $S_i$, a scalar variable, respectively. The microscopic state of the system is then denoted by $\{\vec r_i\}$ and $\{S_i\}$, the set of the degree of freedom of all monomers. 

In general, the number of feasible spin states depends on the model under consideration. Considering an effective model of chromatin, spins generally adopt two or three states. This is a class of Potts model\cite{Potts_Wu} as a magnetic model. When considering this model as a model of chromatin, the spin variable $S_i$ represents the histone modification state of the $i$-th nucleosome. Assuming three states of the spin degrees of freedom, the state with $S_i=0$ is assigned to a non-modified neutral state, and the states with $S_i=\pm 1$ are assigned to different modified states. For example, in heterochromatin, the two main histone modifications are the trimethylation of histone H3 at lysine 9 (H3K9me3) and at lysine 27 (H3K27me3), which correspond to the two modified states. 

The polymer Potts model was first demonstrated to exhibit a simultaneous conformational and magnetic order transition, using molecular dynamics simulations\cite{PhysRevX.6.041047}. Subsequently, it was demonstrated to be a first-order phase transition by both mean-field theory and corresponding molecular dynamics simulations\cite{PhysRevE.100.052410}. This is in contrast to the coil--globule transition of homopolymer without an internal degree of freedom, which is considered to be a second-order phase transition in most theoretical studies\cite{RevModPhys.50.683,doi:10.1021/acs.macromol.7b01518}. The coupling between a three-dimensional structure and the internal degree of freedom in one dimension leads to the remarkable effect of changing the order of the transition. In a slightly different but similar model, a different construction of the free energy of the model also confirms the existence of a first-order phase transition, and it is noted that the jump in the magnetic order parameter at the transition temperature is enhanced by the coupling to the polymer conformation\cite{AdachiKawaguchi2019}.

In the previous studies\cite{PhysRevE.100.052410,AdachiKawaguchi2019,Garel1999} on the polymer Potts model or magnetic polymer model, the effect of the offset of the interaction energies between monomer segments was not been seriously considered. In the absence of conformational degrees of freedom, the interaction energy offset is a shift in the energy origin and does not affect the equilibrium state. However, in the case of the polymer Potts model, the relative relationship between the energy of the polymer conformation and that of the internal degrees of freedom may affect the equilibrium state. This may be why only simultaneous first-order transitions of the conformational and magnetic order were observed previously because the effect of the offset has not been studied extensively. In fact, the effect of the offset in the interaction energy on the modification state of biological systems such as chromatin is noteworthy. In this study, we introduce an offset in the interaction energy between monomer segments into the polymer Potts model and investigate the equilibrium phases of the model using the mean-field approximation and molecular dynamics simulations.

The remainder of this paper is organized as follows: in Sec.~\ref{sec:latticePolymerPotts}, a polymer Potts model of a lattice with spin as an internal degree of freedom on a monomer is introduced. Moreover, its phase diagram is illustrated by a  mean-field analysis. In particular, we demonstrate that by controlling the energy offset, conformational and magnetic order formation can be separated, and a phase with a compact conformation and magnetic disorder (called the compact-disordered phase) would emerge. We also discuss the phase diagram obtained when the modified-state fraction is controlled. In Sec.~\ref{sec:MD}, we present the numerical results of the molecular dynamics simulations for another polymer Potts model introduced as an off-lattice model. For a model with an energy offset different from that in previous research\cite{PhysRevE.100.052410}, we demonstrate that the compact-disordered phase and a similar two-step phase transition identified in the lattice model are observed in the behavior of certain physical quantities as a function of temperature. Finally, Sec.~\ref{sec:summary} presents the summary and discussion.  

\section{Mean-field theory for a lattice polymer Potts model}
\label{sec:latticePolymerPotts}
\subsection{Lattice model}
In this section, the polymer Potts model is defined precisely on a lattice and analyzed with the mean-field theory. 
For a simple lattice polymer model, the configuration of the polymer is limited to be on a simple cubic lattice with a lattice spacing $a$ and is confined in a volume $V$ with a monomer density $\rho=N/V$. Here, only the exclude-volume effect is considered, assuming that the elastic energy of the polymer is omitted. The Hamiltonian of the system consists of the product of the contributions with respect to the configuration $\{\vec r_i\}$ and the spin $\{S_i\}$. It is expressed as  
\begin{align}
H(\{\vec r_i\}, \{S_i\})  =&\frac{1}{2} \sum_{i \neq j} \Delta(\vec r_i, \vec r_j) J(S_i, S_j)\nonumber\\
&- h \sum_i S_i-\mu \sum_i S_i^2, 
\label{eqn:Hamiltonian}
\end{align}
where $h$ is the external field, $\mu$ is the chemical potential that controls the fraction of the modified states, $J(S_i, S_j)$ denotes the magnetic interaction, and $ \Delta(\vec r_i, \vec r_j)$ is the adjacency matrix of the polymer with the lattice spacing $a$. It is expressed as
\begin{equation}
\Delta\left(\vec r_i, \vec r_j\right)= 
\begin{cases}
1, & |\vec r_i-\vec r_j| = a, \\
0, & \text { otherwise. }
\end{cases}
\label{eqn:delta}
\end{equation}

Here we assume that there are three spin states, as in the example explained in the previous section, and set the values of spin to $S_i=-1$, $0$, and $1$. Specifically, we consider $S_i=1$ to represent the modified state H3K9me3 and $S_i=-1$ to represent H3K27me3. This is based on several reports\cite{zhangInterplayHistoneModifications2015} that the two modified states are mutually exclusive. In the case of chromatin, the properties of other molecules responsible for the modified states require consideration when setting the magnetic interactions. 
One is molecules called the ``reader" that specifically recognize the epigenetic modification. The reader molecules for H3K9me3 and H3K27me3 are HP1 and PRC2, respectively. They are also known to bridge between the nucleosomes with the same modification and play the role of effective interactions between nucleosomes\cite{MACHIDA2018385, poepselCryoEMStructuresPRC22018}. In addition, there are other ``writer" molecules that deposit the biochemical modification on nucleosomes, and HP1 is known to recruit the writer molecules of H3K9me3, and one of PRC2 subunits is known to be the writer of H3K27me3 itself\cite{zentnerRegulationNucleosomeDynamics2013, liJarid2PRC2Partners2010, felsenfeldBriefHistoryEpigenetics2014}.

Incorporating the above properties in a simplified form, our model employs the magnetic interaction $J(S_i,S_j)$ between the monomers given by 
$$J\left(S_i, S_j\right)=\left\{\begin{array}{cl}
-\frac{\varepsilon}{2}(c+1) &\quad S_i=S_j= \pm 1, \\
-\frac{\varepsilon}{2}(c-1) &\quad \text {otherwise, }
\end{array}\right.$$
where $c$ is a parameter that provides the offset of the magnetic energy of the system, and the positive constant $\varepsilon$ represents the coupling amplitude. 
This implies that the nucleosomes in this model prefer to be in close proximity independent of their modification state when $c > 1$ and prefer more to have an identical modification state because $\epsilon > 0$.
With the spin variables, the interaction energy function has bilinear and bi-quadratic terms. They are explicitly expressed as 
\begin{equation}
J\left(S_i, S_j\right)=\frac{\varepsilon}{2}\left(1-S_i S_j-S_i^2 S_j^2-c\right). 
\label{eqn:Potts_interaction}
\end{equation}
 In the case of only the spin system without the polymer conformation,  the parameter $c$ yields only a shift in the free energy. However, in the polymer Potts model, the parameter $c$ has a nontrivial effect on the free energy by coupling the conformation and spin degrees of freedom of the polymer through the $\Delta$ term in Eq.~(\ref{eqn:Hamiltonian}). As is explained subsequently, the phase diagram of the system and the order of the phase transition depend explicitly on $c$. This energy term, including the bilinear and biquadratic exchange interactions, has a $Z_2$ symmetry, rather than a $Z_3$ symmetry of the conventional three-state Potts model. It is called the $S=1$ Ising model in the field of magnetism in statistical physics. 

The partition function of the system is then given by 
\begin{align}
   Z(\beta, h, \mu; c)=\sum_{\text {SAW }} \sum_{\vec S} & \exp \left(  -\frac{\beta}{2} \sum_{i, j=1}^N \Delta\left(\vec r_i, \vec r_j\right) J\left(S_i, S_j\right) \right. \nonumber \\
    & \left.+\beta h \sum_i S_i+\beta \mu \sum_i S_i^2\right),
    \label{eqn:PP_PartitionF}
\end{align}
where $\sum_{\rm SAW}$ represents the sum of $\{\vec r_i\}$ over self-avoiding walks for the polymer conformation. The order parameter for the coil--globule transition in this model is the monomer concentration $\rho$. That is, the swollen phase is characterized by $\rho=0$, whereas the compact phase by $\rho>0$. The order parameter for the  magnetic phase transition is the magnetization given by 
\begin{equation}
    m=\frac{1}{N}\sum_i\langle S_i\rangle, 
\end{equation}
and the conjugate quantity to the chemical potential $\mu$ is 
\begin{equation}
    D=\frac{1}{N}\sum_i\langle S_i^2\rangle, 
\end{equation}
where $\langle\cdots\rangle$ denotes the expectation with respect to the equilibrium thermal state. The second of these is the fraction of spins adopting $\pm 1$ or in terms of chromatin, the fraction of nucleosomes in modification states with $S_i=\pm 1$. The magnetically disordered phase and the ordered phase of this model are characterized by $m=0$ and $m \neq 0$, respectively. 

At the limit of $\mu\rightarrow\infty$, the spin states are restricted to $\pm 1$. Thus, $D=1$ because the neutral states are eliminated completely, and the model is reduced to the polymer Ising model\cite{Garel1999, PhysRevE.104.024122, PhysRevE.104.054501, papaleIsingModelSwollen2018}.
In a previous study of the polymer Ising model using mean-field theory\cite{Garel1999}, it was demonstrated that the magnetic phase transition and the coil--globule transition occur simultaneously. In particular, a first-order transition occurs for sufficiently small magnetic fields including zero, and a second-order transition occurs for large fields. Subsequently, Monte Carlo studies on finite-dimensional lattice models\cite{PhysRevE.104.024122,PhysRevE.104.054501,papaleIsingModelSwollen2018} yielded qualitatively consistent results with the mean-field theory concerning the phase diagram for three-dimensional simple cubic lattices. However, these studies indicated that no simultaneous phase transition occurs for two-dimensional square lattices. Nevertheless, the effect of  $c$ in Eq. ~(\ref{eqn:Potts_interaction}) was not considered in previous studies. 

\subsection{Mean-field theory}
Using the mean-field approximation for self-avoiding walks and the dilution approximation for the adjacency matrix $\Delta(\vec r_i, \vec r_j )$ used in the previous study\cite{Garel1999}, the free-energy density $f(\beta,h,\mu;c)$ is obtained as 
\begin{align}
\beta f(\beta,h,\mu; c) &= \min_{\phi,\psi, \rho}\frac{1}{\rho \alpha}\left(\phi^2+\psi^2\right) \nonumber\\&-\ln \left\{1+2 e^{\psi+\beta\mu} \cosh (\phi+\beta h)\right\} \nonumber\\
&+\frac{1-\rho}{\rho} \ln (1-\rho)+\frac{\rho \alpha}{4}(1-c)-\ln \left(\frac{q}{e}\right),
\label{eqn:FE_PP}
\end{align}
where $q$ is the coordination number of the lattice defined by the model and $\alpha=\beta\varepsilon q$. The extremum conditions for the auxiliary fields $\phi$ and $\psi$ yield the following saddle-point equations: 
\begin{align}
    \phi &= \frac{\rho\alpha}{2} \frac{2e^{\psi+\beta \mu} \sinh (\phi + \beta h)}{1+2 e^{\psi+\beta \mu} \cosh (\phi + \beta h)},\label{eqn:SPE_1}\\
    \psi &= \frac{\rho\alpha}{2} \frac{2 e^{\psi+\beta \mu} \cosh (\phi + \beta h)}{1+2 e^{\psi+\beta \mu} \cosh (\phi + \beta h)},
    \label{eqn:SPE_2}
\end{align}
respectively. 
Details of the derivation of the free-energy density are provided in Appendix.~\ref{sec:details_MFT}. 

\begin{figure}[ht]
\includegraphics[ width=0.9\linewidth]{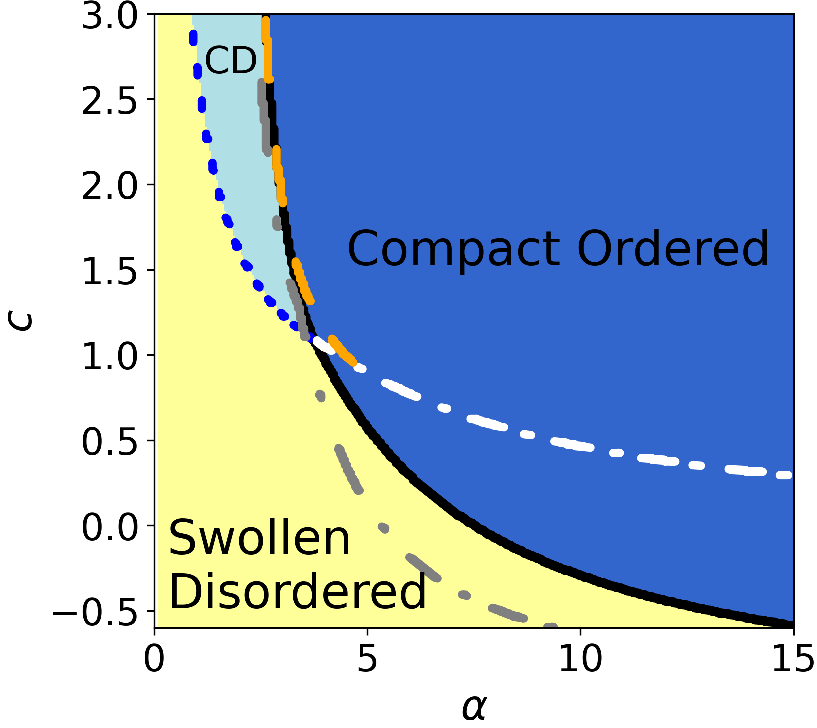}
\caption{\label{fig:epsart} (Online color) Phase diagram of the lattice polymer Potts model in the $\alpha$ -- $c$ plane at $\mu=0$ and $h=0$. 
$m=0$ and $\rho=0$ in the swollen-disordered phase,  and $m\neq0$ and $\rho>0$ in the compact-ordered phase. CD stands for the compact-disordered phase, which is characterized by $m=0$ and $\rho>0$. 
The solid and dotted lines indicate the first-order and second-order phase transitions, respectively. The broken and dashed lines in the compact-ordered phase represent the instability line of the compact-disordered and swollen-disordered solutions, respectively.  The broken line in the swollen disordered phase represents the instability line of the compact ordered solution.}
\label{fig:1}
\end{figure}

Using the solutions $\phi^*$ and $\psi^*$ of the saddle-point equations, the order parameter $m$ and the fraction of the modified states $D$ are expressed as follows: 
\begin{align}
    m & = \frac{2}{\rho\alpha}\phi^*, \\
    D & = \frac{2}{\rho\alpha}\psi^*,
\end{align}
respectively. The thermodynamically stable monomer density $\rho^*$ is determined from the extremum condition in Eq.~(\ref{eqn:FE_PP}) for $\rho$. The phase diagram of the mean-field approximation of this model can be obtained by numerically solving the saddle-point equations. Fig.~\ref{fig:1} shows the phase diagram of the polymer Potts model in the $\alpha-c$ plane at $\mu=0$ and $h=0$. There are three phases: a swollen-disordered (SD) phase with $\rho = 0$ and $m = 0$, a compact-disordered (CD) phase with $\rho > 0$ and $m = 0$, and a compact-ordered (CO) phase with $\rho > 0$ and $m \neq 0$. Another possible phase is the swollen-ordered phase with $m\neq0$ and $\rho=0$, which cannot exist as an equilibrium state in the mean-field analysis because of the saddle-point equations of Eqs.~(\ref{eqn:SPE_1}) and (\ref{eqn:SPE_2}), while a previous study suggests that it appears under nonequilibrium\cite{genomic-bookmarking}. 

First, we verify the phase structure at the two limits, $\mu\rightarrow\infty$ and $\mu\rightarrow -\infty$. The polymer Ising model studied in the previous work\cite{Garel1999} corresponds to a system at $\mu\rightarrow\infty$, in particular, fixed at $c=0$. Our results provide an extended phase diagram with respect to the parameter $c$, which is shown in Appendix.~\ref{sec:pD_polymerIsing}. It should be emphasized that in the region where $c$ is small, including $c=0$, there is a first-order phase transition between the SD and the CO phases, but the CD phase is present between the two phases as $c$ is increased. In the opposite limit, $\mu\rightarrow-\infty$, the CO phase is absent because the spins are only neutral states with $S_i=0$. This indicates a coil--globule transition with no spin degrees of freedom. The phase transition is second-order, and the phase boundary of the second-order phase transition is obtained as $c_c(\alpha)=1+\frac{2}{\alpha}$. 

\begin{figure}[t]
    \centering
   \includegraphics[width=\linewidth]{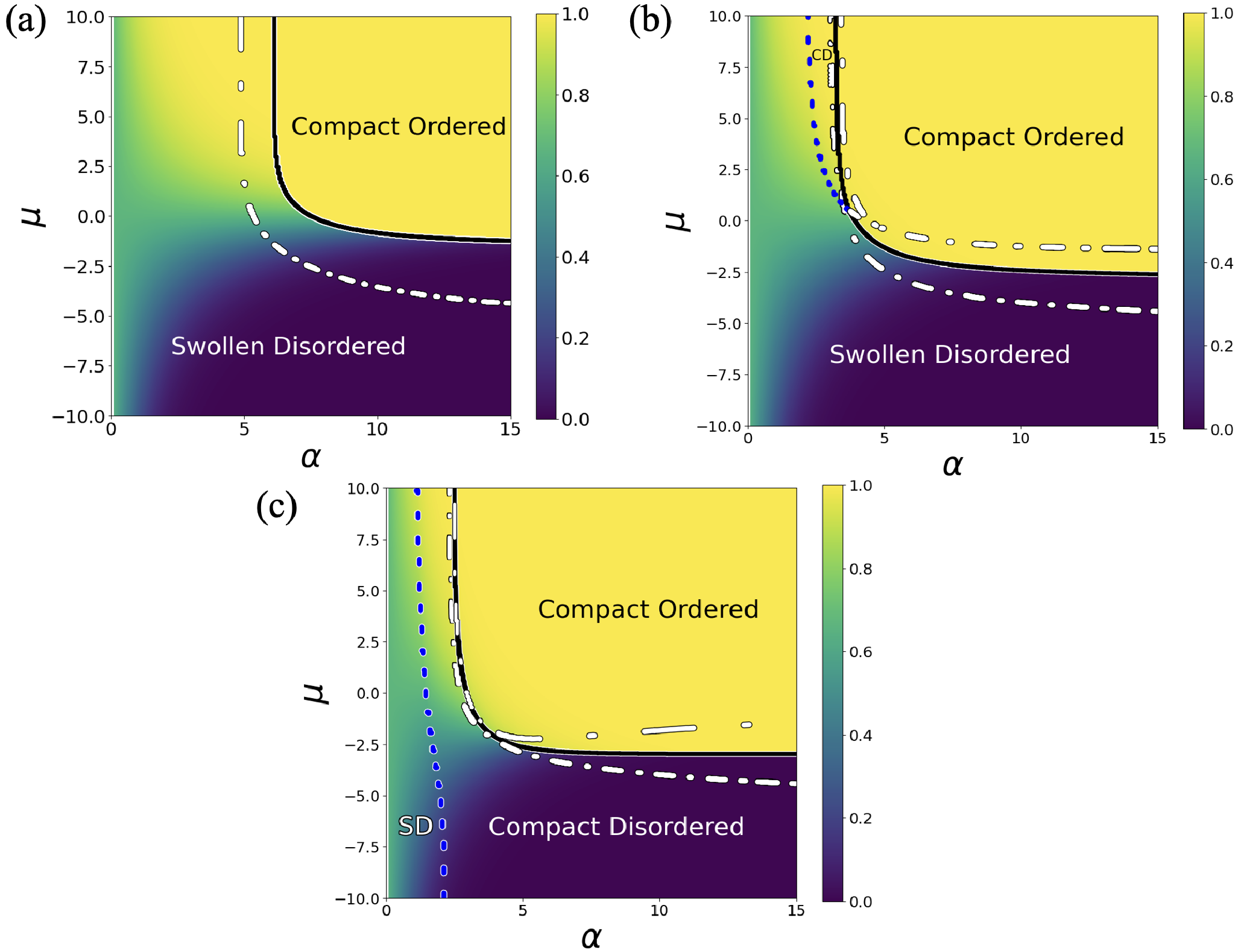}
    \caption{(Online color) Phase diagrams of the lattice polymer Potts model in the $\alpha$ -- $\mu$ plane at (a) $c=0$, (b) $c=1$ and (c) $c=2$. The solid and dotted lines indicate the first-order and second-order phase transitions, respectively. CD stands for compact-ordered phase. The dotted dashed white lines represent the instability conditions. The heat map in the figures represents the average ratio of modified states $D$.}
    \label{fig:PD-amu}
\end{figure}

Next, we present the phase diagram in the parameter region between the two limits of $\mu$. Fig.~\ref{fig:PD-amu} shows the phase diagram in the $\alpha-\mu$ plane, given three specific values of $c$. Considering that $c$, which appears only in the linear term of $\rho$ in the free energy of Eq.~(\ref{eqn:FE_PP}), contributes to stabilizing the SD phase, it is reasonable that only the SD and CO phases are stable at small $c$ values and that the CD phase does not appear, as shown in Fig.~\ref{fig:PD-amu}(a). The phase diagram obtained in most previous studies is basically this phase structure because the parameter $c$ is not taken into account. However, as shown in Fig.~\ref{fig:PD-amu}(b) and (c), the region where the CD phase is thermodynamically stable increases with an increase in $c$. For example, in Fig.~\ref{fig:PD-amu}(c), for a fixed temperature parameter $\alpha$, changing the chemical potential $\mu$, (i.e., increasing the number of modified monomers) results in a first-order phase transition from the CD phase to the CO phase over a wide range of $\alpha$.

\begin{figure}[t]
    \includegraphics[width=0.8\linewidth]{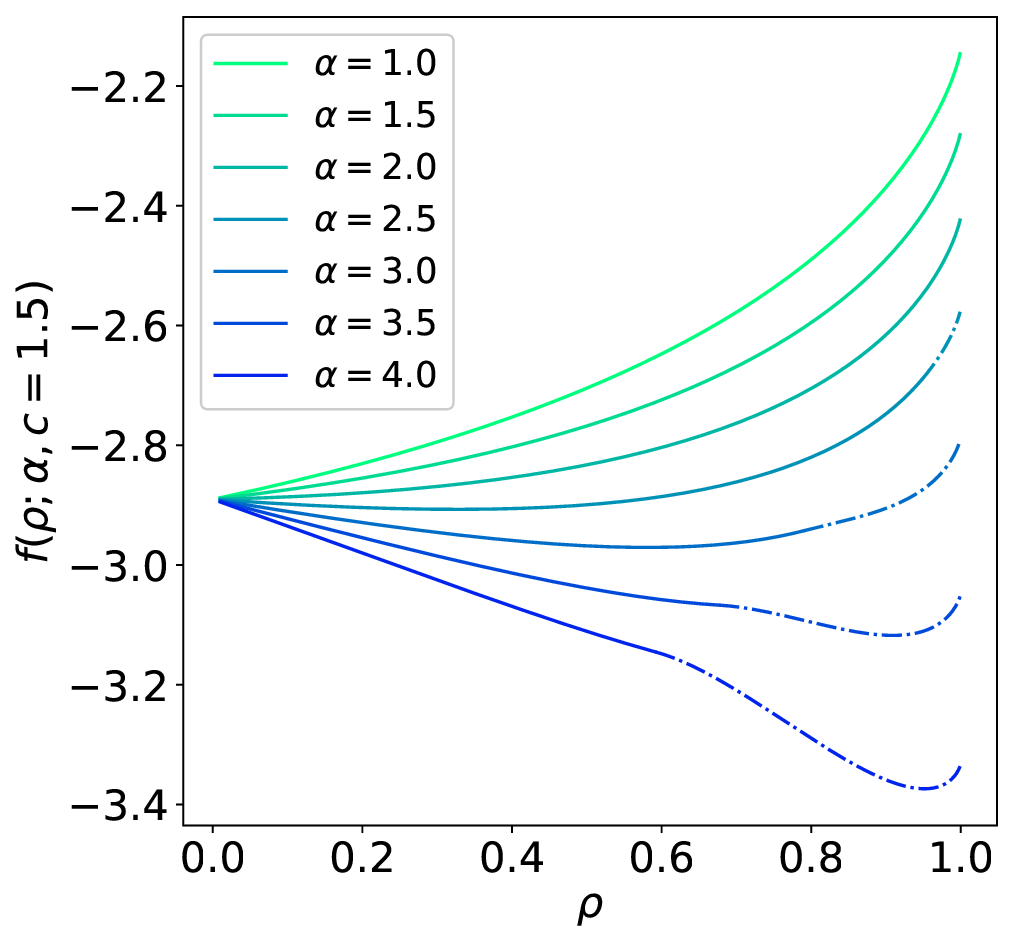}
    \caption{(Online color) Free-energy landscape as a function of the density $\rho$ for $c=1.5$ and certain values of $\alpha$. The solid and dashed lines represent the free-energy densities corresponding to the saddle-point equations with $\phi = 0$ and $\phi \neq 0$, respectively.}
    \label{fig:FE}
\end{figure}

To understand the mechanism of the phase transition in more detail, we evaluated the free-energy landscape. It is defined as a function of the density $\rho$ by taking only the $\phi$ and $\psi$ of the extreme values on the right-hand side of Eq.~(\ref{eqn:FE_PP}). Fig.~\ref{fig:FE} illustrates the density dependence of the free-energy landscape on several typical parameters. The free-energy density is characterized by the solutions of the saddle-point equations with $\phi=0$ and $\phi>0$. At sufficiently high temperatures (i.e., small $\alpha$), only solutions with $\phi=0$ exist, and the free-energy landscape has a minimum at $\rho=0$, which corresponds to the SD phase. As the temperature decreases, this SD phase solution becomes unstable while maintaining $\phi=0$, and the minimum shifts to $\rho>0$. This implies a second-order phase transition from the SD phase to the CD phase. At lower temperatures, a minimal solution with $\phi>0$ appears as another branch of the saddle-point equations. Eventually, this local minimal solution takes a lower value of the free-energy density than that of the CD phase, which is the phase transition from the CD phase to the CO phase. This implies a first-order phase transition in which there is a discontinuous jump in the density and the magnetic order parameter. This crossing of the free energies of the CD and CO phases is a characteristic of the first-order phase transition, which is the mechanism of the phase transition between these two phases in the mean-field theory of this model. 

This first-order phase transition mechanism indicates the existence of the CD phase as a metastable state.
Notably, as shown in Fig.~\ref{fig:1}, the instability line of the metastable state of the CD phase is obtained explicitly using the mean-field theory. This indicates that the CD phase could be dynamically observed as a metastable state in the CO phase. However, the metastable state exists in an extremely small region for large $c$, while the metastable state of the SD phase is rather extended in the CO phase. This implies that the metastable solution of the CD phase with $\rho>0$ and $\phi=0$ destabilizes in the $\phi$ direction immediately after entering the CO phase.  This is an important insight that we have obtained by introducing $c$ and the full mean-field theory.   

\section{Molecular Dynamics Simulation for an off-lattice model}
\label{sec:MD}
This section discusses a slightly realistic three-dimensional off-lattice polymer Potts model for chromatin. The lattice model discussed in the previous section is a sort of effective model. Its mean-field theory provides many insights into the three-dimensional polymer Potts model. To be specific, it was indicated that the three phases, SD, CD, and CO phases, are thermodynamically stable, and a phase diagram has been derived. However, in general, it is considered that the mean-field theory does not correctly take into account the effect of fluctuations in finite dimensions. In particular, the mean-field predictions of the order of transitions are often modified in finite dimensions. Therefore, we perform molecular dynamics (MD) simulations to directly investigate the behavior of the three-dimensional off-lattice polymer Potts model with three spin states as in the previous section. 

\subsection{Off-lattice polymer Potts model}
In our polymer Potts model, a polymer chain consisting of $N$ monomer segments has two types of potential energies: the bonding energy between neighboring monomers along the polymer chain and the non-bonding energy between monomers physically proximate to each other. The second of these depends on the internal degrees of freedom, which is a feature of the polymer Potts model.  
Following 
the Kremer-Grest\cite{KremerGrest} polymer model, the bonding energy is assumed to be only a function of the distance $r$ between the adjacent monomer segments and to follow the Finitely Extensible Nonlinear Elastic\cite{KremerGrest} (FENE) potential given by 
\begin{equation}
U^{\mathrm{FENE}}(r)= \begin{cases}
-\frac{1}{2} k R_{0}^{2} \ln \left(1-\left(\frac{r}{R_{0}}\right)^{2}\right) & \mbox{for } r \leq R_{0}, \\
\infty & \mbox{for } r>R_{0}, 
\end{cases}
\end{equation}
where $R_0$ is the length scale and $k$ is the coupling constant.

For non-bonding interactions between the monomer segments, we use a potential acting on the monomers at distances within a finite cutoff $r_c$, which depends on the internal state of the monomer segments. Specifically, it is described by the shifted and truncated Lennard-Jones (LJ) potential defined as 
\begin{equation}
U_{\mathrm{LJ}}^{'}(r)=
\begin{cases} U_{\mathrm{LJ}}(r) - U_{\mathrm{LJ}}(r_c)
& r<r_c  \\ 0 & r \geq r_c \end{cases},
\end{equation}
where $ U_{\mathrm{LJ}}(r)$ is the Lennard-Jones potential given by  
\begin{equation}
U_{\mathrm{LJ}}(r)= A \left[ 4\varepsilon_\mathrm{LJ}\left\{\left(\frac{\sigma}{r}\right)^{12}-\left(\frac{\sigma}{r}\right)^{6}\right\}\right],
\end{equation}
where $A$ is a dimensionless parameter for the potential magnitude, and $\varepsilon$ and $\sigma$ are the units of the energy and length scales, respectively.  
When we set $r_c/\sigma = 2^{1/6}$, this interaction potential is purely repulsive as a special case, which is called Weeks-Chandler-Andersen\cite{WCA} (WCA) potential. For the other LJ potentials, the cutoff $r_c$ is set to be $r_c/\sigma=1.8$ in the simulations. 

\begin{figure}[t]
    \centering
  \includegraphics[width=\linewidth]{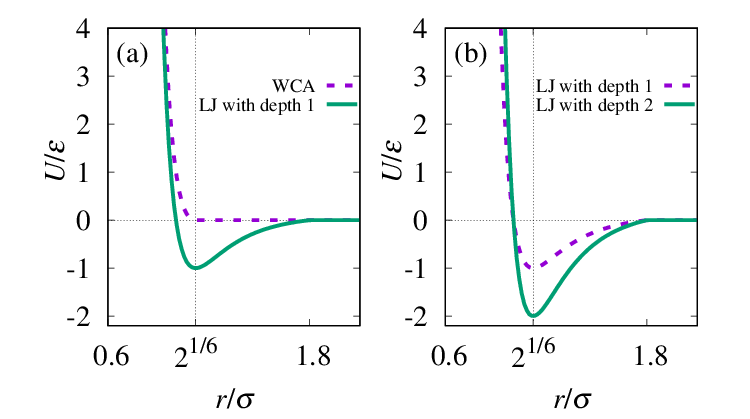}
    \caption{(Online Color) Interaction potentials used in our simulations. The solid lines represent a pair potential when the monomer spins are $\pm 1$ and aligned, and the dashed lines represent a pair potential when these are not. (a) Model 1: The solid line is the shifted-truncated LJ potential with a depth of $1$, and the dashed line is the WCA potential. (b) Model 2: Both solid and dashed lines are shifted-truncated LJ potentials with depths $1$ and $2$, respectively. The cutoff $r_c$ of the LJ potentials is set to be $r_c/\sigma=1.8$. }
    \label{fig:MD-interactions}
\end{figure}

To study the role of the energy offset revealed in the analysis of the lattice model in the previous section, we consider the non-bonding interactions that change the energy offset while maintaining the amplitude of the energy gain by aligning the monomer modification states fixed. While the energy gain in the lattice model can be explicitly given, it is not necessarily apparent in the off-lattice model. Here, the energy gain is assumed to be the difference between the energy minima of the two potentials. In our simulations, for the case of a small offset, as shown in the left panel of Fig.~\ref{fig:MD-interactions}, the non-bonding interaction potential is the LJ potential with a depth of $1\textrm{k}_{\textrm{B}}T$ when the nucleosome pairs are both modified and in an identical state, and the interaction potential is the WCA potential when the nucleosome pairs are in different states or both are in a neutral state. Thus, the energy gain is $1\textrm{k}_{\textrm{B}}T$. Hereafter, we will refer to the model as Model 1. When the offset is large, as shown in the right panel of Fig.~\ref{fig:MD-interactions} and referred to as Model 2, the potential is the LJ potential with depths of $2\textrm{k}_{\textrm{B}}T$ and $1\textrm{k}_{\textrm{B}}T$ when the pairs of nucleosomes are in an identical modified state and in all other pairs of modification states, respectively. Again, the energy gain is $1\textrm{k}_{\textrm{B}}T$. 

In our MD simulations, the monomer dynamics in the polymer Potts model followed a Langevin equation. We used the Lennard-Jones dimensionless reduced unit in which $m$, $\sigma$, $\varepsilon_\mathrm{LJ}$, and  $\text{k}_{\text{B}}$ were set to unity. The time integrator of the molecular dynamics method is a velocity-Verlet algorithm with a time step of $0.005\tau$. Here $\tau$ is the time unit defined as $\tau=\sqrt{{m\sigma^2}/{\varepsilon_\mathrm{LJ}}}$.  The Langevin dynamics simulations 
were performed using a Large-scale Atomic/Molecular Massively Parallel Simulator (LAMMPS)\cite{LAMMPS}.
The spin degrees of freedom of each monomer were updated using a Monte Carlo method during the MD simulation. This Monte Carlo update is performed using a heat-bath-type transition probability and is attempted at every $10^3$ time step of the MD by the number of monomers. The chemical potentials of all the modification states were set to zero in our simulations. 
In typical simulations, unless otherwise specified, the linear dimension of the simulation box is $L=100$, and the length of the polymer is $N=500$. The box size $L$ is considered sufficiently large compared with the gyration radius of the coil state of a polymer of length $N$. 

The pre-equilibration process of the system is described, before showing the results of the main simulations. 
The initial polymer conformation is set as a freely jointed polymer. The bond length is set to minimize the FENE potential for the initial condition, and the system evolves under the soft repulsive interaction between the monomers. This is given by  
\begin{equation}
E=A_\mathrm{soft}\left[1+\cos \left(\frac{\pi r}{r_c}\right)\right] \quad r<r_c,
\end{equation}
where $A_\mathrm{soft}$ denotes the magnitude of the soft potential. In the pre-equilibration process, $A_\mathrm{soft}$ is first increased linearly from $0$ to $30$ in the initial $10^3$ time steps to remove the overlap between the monomers. Subsequently, the soft repulsive interactions are replaced by the LJ potentials described above, and the system evolves $10^7$ time steps to attain the equilibrium state. To verify that the system has attained the equilibrium state, this pre-equilibration process is performed for several polymer sizes to examine whether the Flory scaling of the gyration radius in the coil state holds. The final state of the pre-equilibration process at a sufficiently high temperature is used as the initial condition for the subsequent main simulation.

\subsection{Results of MD simulations}
\begin{figure}[t]
   \includegraphics[width=0.8\linewidth]{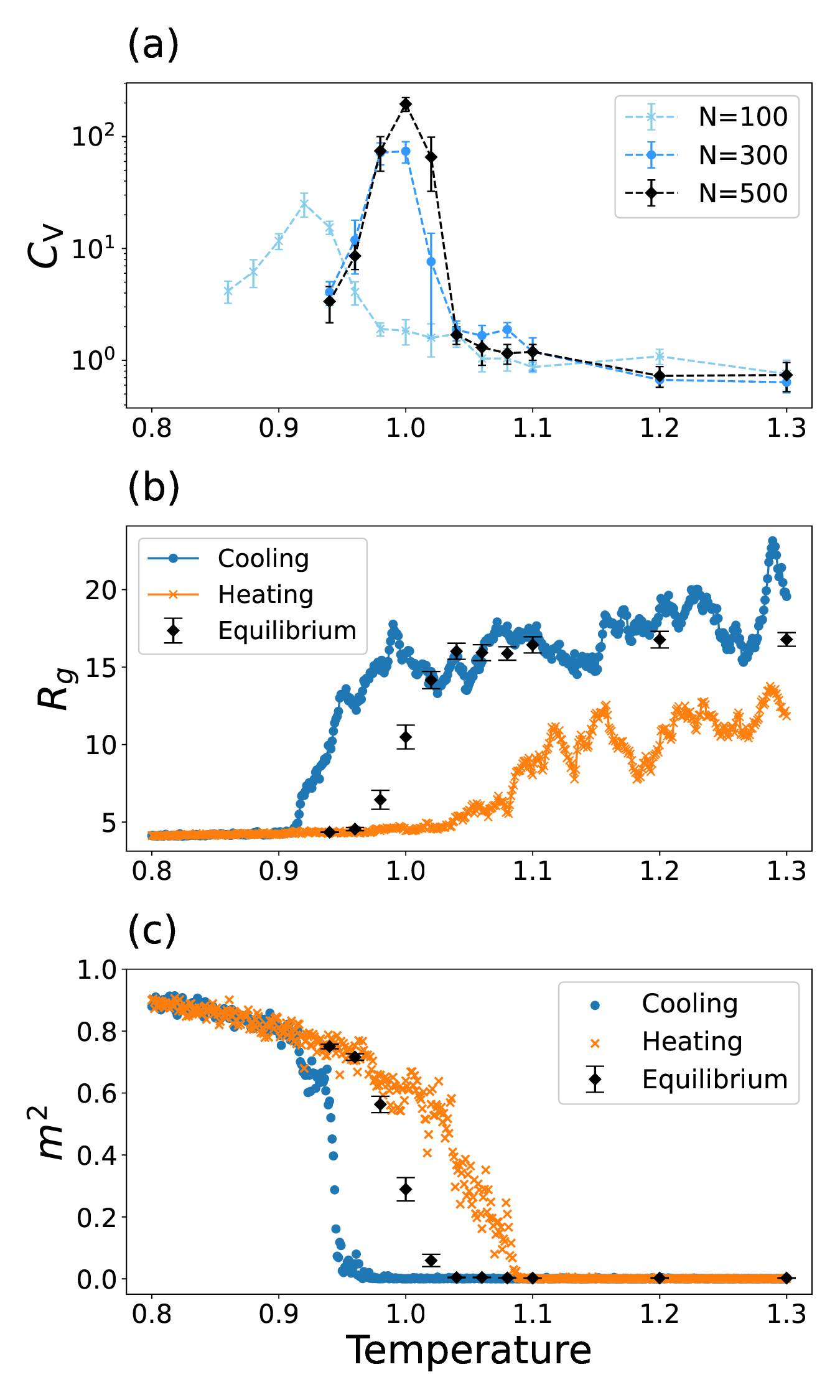}
    \caption{(Online Color) Temperature dependence of (a) specific heat $C_V$, (b) gyration radius $R_g$, and (c) squared magnetization $m^2$ in Model 1 with the WCA and LJ potentials with a depth of $1$ in equilibrium by black marks. The results for $R_g$ and $m^2$ under non-equilibrium conditions are also shown during the heating (crosses) and cooling (circles) processes. }
    \label{fig:smallC}
\end{figure}

First, we present the MD simulation results of Model 1. Herein, the interaction potential between monomer segments of different states and between segments of the neutral state $S_i=0$ and those of other states is the WCA potential, and the interaction between segments in the same modification states with $S_i = 1$ or $-1$ is the attractive LJ potential with depth unity in the LJ reduced unit. A first-order-like transition between the SD and CO phases was reported in a previous MD simulation study\cite{PhysRevX.6.041047} of a different but nearly identical model with different bonding potentials. We also verified that a first-order-like transition occurs, as shown in Fig.~\ref{fig:smallC}, from the temperature dependence of the specific heat, equilibrium thermal averages of the gyration radius $R_g$, and those of the squared magnetization $m^2$, together with the results of non-equilibrium simulations during heating and cooling processes. 

Equilibrium simulations were performed for 50 independent configurations equilibrated at $T=1.3$ at a cooling rate of 0.01 every $1.5 \times 10^6$ time steps. Thermal equilibrium average values were obtained from the weighted averages over the process using the annealed importance sampling method\cite{nealAnnealedImportanceSampling2001}.  In the nonequilibrium simulations, the polymer conformation was first equilibrated at $T=1.3$. Moreover, the temperature was reduced by $0.001$ every $10^4$ time steps to the lowest temperature $T=0.8$ and increased from $T=0.8$ to $T=1.3$ subsequently at an equal rate. The short-term averages of $R_g$ and $m^2$ at each temperature during this cooling and heating process are shown in Fig.~\ref{fig:smallC} as ``Cooling" and ``Heating", respectively. 

As the temperature decreased, the specific heat exhibited a single sharp peak at a certain temperature around which the gyration radius decreased substantially and the magnetization began to adopt a finite value. In addition, a remarkable hysteresis in $R_g$ and $m^2$ was observed near the temperature at which the specific heat was maximized. These are characteristics of first-order phase transitions. Meanwhile, there was no clear evidence of the CD phase in our MD simulations. This indicated a single phase transition from the SD to the CO phases as the temperature decreased. This, in turn, indicated that the interaction potentials in this simulation corresponded to the smaller $c$ regime in the lattice model discussed in previous sections. 

\begin{figure}[t]
    \centering
   \includegraphics[width=0.8\linewidth]{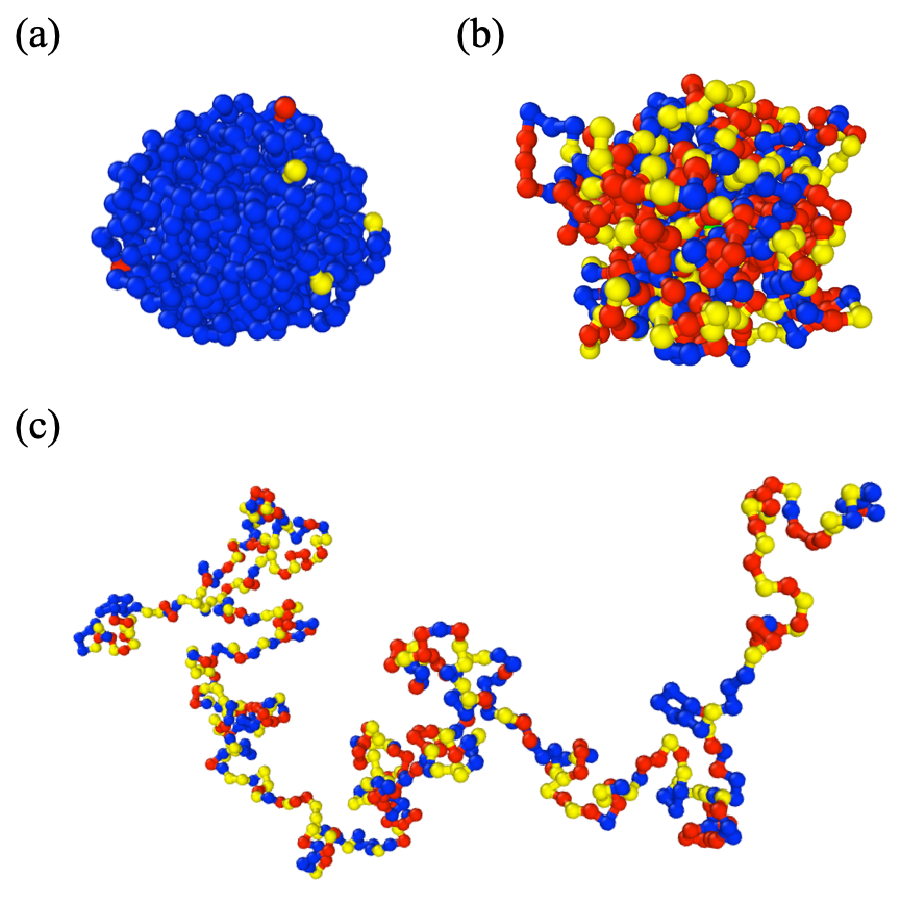}
    \caption{(Online color) Snapshots of the MD simulations at each temperature at (a) $T=1.0$ (b) $2.0$, and (c) $3.0$.  
    The yellow monomers represent the neutral state, and the blue and red ones represent the modified states. The simulation outputs were visualized with OVITO software\cite{Stukowski_2010}
    }
    \label{fig:3phases-md}
\end{figure}

Next, we discuss the simulation results of another model, Model 2. Here, the interaction potentials between monomer segments in different states and/or the neutral state $S_i= 0$, and those between segments in the same modified state with $S_i = 1$ or $-1$ are the attractive LJ potentials with depths of $1$ and $2$ in the LJ reduced unit, respectively. Similar to the previous model, this model switches between two potentials depending on the internal degrees of freedom of the monomer segments. The energy minima are reduced, whereas the difference between the minima of the two potentials is fixed at $1$. We consider this as equivalent to increasing the offset parameter $c$ in the lattice model described in the previous section. Three characteristic snapshots of the model MD simulations at different temperatures are shown in Fig.~\ref{fig:3phases-md}. As shown in the figure, there is a typical compact conformation with the internal states of the polymer remaining disordered. This indicates the existence of a CD phase in the intermediate temperature regime in addition to the SD and CO phases. 

\begin{figure}
    \centering
   \includegraphics[width=0.8\linewidth]{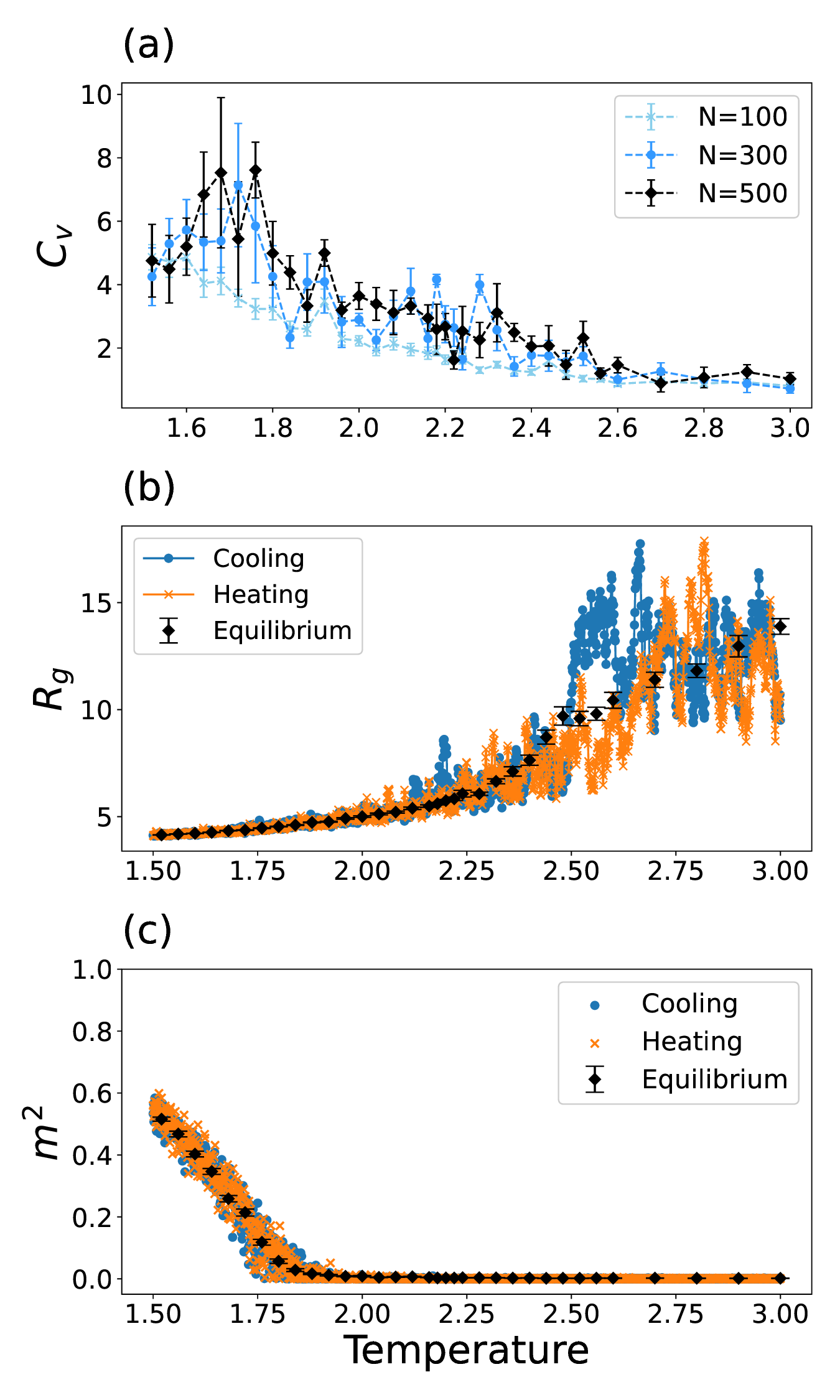}
    \caption{ Temperature dependence of (a) specific heat $C_V$, (b) gyration radius $R_g$, and (c) squared magnetization $m^2$ in Model 2 in equilibrium by black marks. The results for $R_g$ and $m^2$ under non-equilibrium conditions are also shown during the heating (crosses) and cooling (circles) processes.}
    \label{fig:hysteresis_large_c}
\end{figure}

To investigate the phase structure of this model in more detail, the equilibrium physical quantities are calculated from $50$ runs of MD simulations for $N=300$ and $500$, and $300$ runs for $N=100$. Fig.~\ref{fig:hysteresis_large_c} shows the temperature dependence of the specific heat and thermal expectation values of the gyration radius and magnetization. It also displays the non-equilibrium simulation, wherein the temperature is reduced from $T=3.0$ to $T=1.5$ and increased to $T=3.0$ at a rate equal to that for Model 1. The figure shows a significant difference between the temperature at which the gyration radius decreases and that at which the magnetization begins to rise. This indicates that the two transition temperatures are different. Furthermore, no difference exists between the observed values during the cooling and heating processes. That is, no significant hysteresis is observed. This implies that both the transitions are of the second order. Although the mean-field analysis of the lattice model predicted that the phase transition on the low-temperature side would be a first-order transition, this MD simulation did not display a tendency for a first-order transition. For hysteresis to be observed, the metastable state must be locally stable in the low-temperature phase, but the mean-field analysis clarified that the destabilization temperature of the metastable state is marginally close to the transition temperature. Thus, it remains possible that the hysteresis is not observed even when a first-order transition occurs. Although the order of the transitions remains to be investigated, the existence of two transitions and a CD phase between them is strongly indicated.    

\section{summary and discussion}
\label{sec:summary}
In this work, we studied the equilibrium phases and phase transitions of the polymer Potts models, both on and off the lattice, as a model of chromatin. In these models, the Potts spin was introduced into each monomer segment of the polymer chain as an internal degree of freedom. It was assumed to adopt three states: two modified states and a neutral state. In the lattice polymer model, we considered the ferromagnetic interactions between the nearest-neighbor monomer segments. We also introduced an offset $c$ to the interaction energy and the chemical potential $\mu$ conjugate to the fraction of the modified segments.  
The analysis based on the mean-field approximation revealed the existence of three equilibrium phases: compact-ordered, compact-disordered, and swollen-disordered. These depend significantly on parameters such as the temperature, $c$, and $\mu$. It also revealed that the phase transition between the compact-ordered phase and other phases is a first-order transition, whereas that between the compact-disordered and swollen-disordered phases is a second-order transition. 

On the other hand, the off-lattice polymer Potts model has been studied by molecular dynamics simulations. The model includes the energy gain by ferromagnetic interactions between adjacent monomer segments using a Potts-spin-dependent potential, as well as an effect corresponding to the energy offset. Our simulation results indicate that the three phases observed in the lattice model also exist in the off-lattice model. In particular, the compact disordered phase distinctly appears only in the intermediate temperature range with large energy offsets. This is qualitatively consistent with the mean-field theoretical predictions for the lattice model.

We observed that a compact-disordered phase is common in both lattice and off-lattice models. The existence of this phase implies the separation of the coil--globule transition for the polymer conformation and the spin order transition for the spin degree of freedom. This study clarified that the condition for its existence is that the energy offset should be large. One previous study\cite{AdachiKawaguchi2019} of the polymer Potts model pointed out that a compact disordered phase can appear when a second virial coefficient of the free-energy model is varied, but in most cases\cite{Garel1999, PhysRevX.6.041047}, only a direct phase transition from swollen-disordered to compact-ordered phases (i.e., a simultaneous phase transition of conformation and spin) appears. This corresponds to a small offset in our model.  

Finally, we discuss the biological implications of the results. The taxonomy of chromatin is described briefly in the Introduction. It has undergone substantial progress. Heterochromatin is classified into constitutive heterochromatin and facultative heterochromatin based on its properties\cite{TROJER20071}. Constitutive heterochromatin is mainly marked by H3K9me3 in gene-poor regions. Meanwhile, facultative heterochromatin is mainly marked by H3K27me3 in cell type-specific gene-rich regions\cite{TROJER20071,zhangInterplayHistoneModifications2015,poonpermFormationMultiLayered2021}. Our model assumes that these H3K9me3- and H3K27me3-modified states are assigned to the spin states. The two macroscopically stable states with positive and negative values of $m$ in the CO phase can be interpreted as corresponding to the states of the constitutive and facultative heterochromatin.  
The CD phase observed in this study may be interpreted as a compacted chromatin region in which the two modifications were mixed.
Recent studies have suggested that other mechanisms of chromatin compaction may occur independently of H3K9me3 or H3K27me3\cite{NICETTO20191}. Further studies from the perspective of mathematical modeling are required.

\begin{acknowledgments}
We are grateful to Ichiro Hiratani for useful discussions.
This work was supported by JST Grant Number JPMJPF2221 and JSPS KAKENHI Grant Number 23H01095.
One of the authors, RN, was supported by the WINGS-FMSP program at the University of Tokyo.
\end{acknowledgments}

\bibliographystyle{apsrev4-2}
\bibliography{PolymerPotts.bib}
\appendix
\section{Details of mean-field calculations}
\label{sec:details_MFT}
The derivation of the free-energy density shown in Eq.~(\ref{eqn:FE_PP}) is explained somewhat more carefully here, while the basic concepts for the approximation and computation closely follow those in Refs\cite{Garel1999}. Suppose $\vec S^{\top}=(S_1,\cdots,S_N)$ and $\vec T^{\top}=(S_1^2,\cdots,S_N^2)$.  Then, the exponential part of the partition function of Eq.~(\ref{eqn:PP_PartitionF}) can be expressed as 
\begin{align}
&\exp \left(-\frac{\beta}{2} \sum_{i, j} \Delta\left(\vec{r}_i, \vec{r}_j\right)  \frac{\varepsilon}{2}\left(1-S_i S_j-S_i^2 S_j^2-c\right)\right) \nonumber\\
=&\exp \left(\frac{\beta\varepsilon}{4} \vec{S}^{\top}\Delta \vec{S}+\frac{\beta\varepsilon}{4}\vec{T}^{\top} \Delta\vec{T}
-\frac{\beta\varepsilon}{4}(1-c)\sum_{i,j}\Delta_{i,j}\right).\nonumber
\end{align}
By introducing the auxiliary fields $\phi$ for $\vec{S}$ and $\psi$ for $\vec{T}$ with the Hubbard--Stratonovich transformation and assuming homogeneity in those fields, we obtain 
\begin{align}
\exp \left(\frac{1}{2} \vec{S}^{\top}\left(\frac{\beta \varepsilon}{2} \Delta\right) \vec{S}\right) &=\int d \phi \exp \left(-\frac{N}{\beta\varepsilon} \phi^2\sum_{i,j}\left(\Delta^{-1}\right)_{ij}\right. \nonumber \\
&+\left.\phi \sum_iS_i\right),\nonumber 
\end{align}
and 
\begin{align}
\exp \left(\frac{1}{2} \vec{T}^{\top}\left(\frac{\beta \varepsilon}{2} \Delta\right) \vec{T}\right) &=\int d \psi \exp \left(-\frac{N}{\beta\varepsilon} \psi^2\sum_{i,j}\left(\Delta^{-1}\right)_{ij}\right.\nonumber\\
&+\left.\psi \sum_iS_i^2\right),\nonumber
\end{align}
The characteristics of the adjacent matrix $\Delta$ of the self-avoiding random walks are generally difficult to determine. However, assuming a compact Hamiltonian path, a mean-field analysis\cite{OID1985} revealed that the sum of all the matrix elements of $\Delta$ and its inverse $\Delta^{-1}$ are $\sum_{ij}\Delta(\vec r_i, \vec r_j)=Nq$ and   $\sum_{ij}(\Delta)^{-1}_{ij}=N/q$, respectively. Here, $q$ is the coordination number. It has been demonstrated to be exact in the special case in a two-dimensional lattice\cite{SuZuki1988}. Furthermore, in the calculation, the dilution effect was approximated as\cite{Garel1999} 
\begin{equation}
    \sum_{ij}\left(\Delta^{-1}\right)_{ij}=\frac{N}{q\rho}.
\end{equation}

By combining these, the sum of the spin degrees of freedom can be obtained. Moreover, the partition function can be obtained as follows:  
\begin{align}
    Z &=\int d\phi d\psi \exp \left[N \left\{-\frac{1}{\rho \alpha}\left(\phi^2+\psi^2\right)\right.\right. \nonumber\\
& \left.\left.+\ln \left(1+2 e^{\psi+\beta \mu} \cosh {(\phi+\beta h)}\right)+\frac{\rho \alpha}{4}(c-1)\right\}\right]Z_\text{SAW},
\label{eqn:Z_MFA}
\end{align}
where $Z_\text{SAW}$ is the partition function for the polymer configuration that represents the number of feasible self-avoiding walks. For this self-avoiding walk, the mean-field approximation yields 
\begin{equation}
    Z_\text{SAW}=\int d\rho \left(\frac{q}{e}\right)^N \exp (-V(1-\rho) \ln (1-\rho)).
\end{equation}
Substituting this into Eq.~(\ref{eqn:Z_MFA}) and evaluating the integral using the saddle-point method, the free-energy density is derived as Eq.~(\ref{eqn:FE_PP}). 

\section{phase diagram of the polymer Ising model}
\label{sec:pD_polymerIsing}
\begin{figure}[h]
\includegraphics[ width=0.9\linewidth]{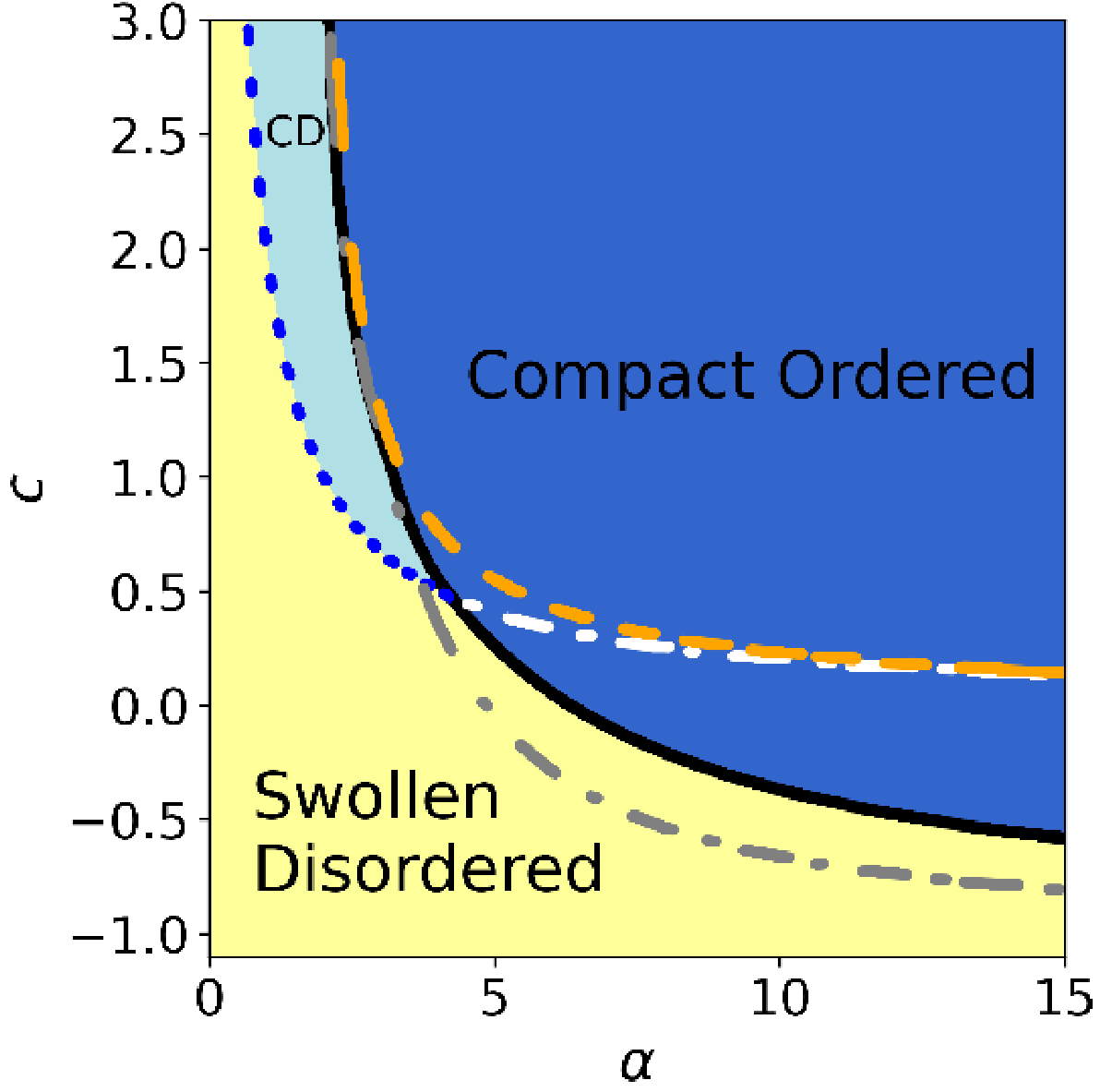}
\caption{(Online color) Phase diagram of the lattice polymer Ising model in the $\alpha$ -- $c$ plane at $h=0$. $m=0$ and $\rho=0$ in the swollen-disordered phase,  and $m\neq0$ and $\rho>0$ in the compact-ordered phase. CD stands for compact-disordered phase, which is characterized by $m=0$ and $\rho>0$. 
The dotted and solid line shows a second-order and first-order transition, respectively.
The dotted dashed lines represent instability conditions and the broken line in the CO phase is the instability condition of the CD solution. 
}
\label{fig:polymerising}
\end{figure}

Our lattice polymer Potts model is reduced to the polymer Ising model by setting the chemical potential $\mu$ controlling the modified states to infinity. Our lattice model of Eq.~(\ref{eqn:Hamiltonian}), (\ref{eqn:delta}) and (\ref{eqn:Potts_interaction}) includes an offset parameter $c$ and the model of $c=0$ is the same as in the previous study\cite{Garel1999}. The free energy of the model including $c$ is obtained by the mean-field theory\cite{Garel1999}.
The phase diagram is shown in  Fig.~\ref{fig:polymerising}. The second-order transition line between SO and CD and the instability condition of SD can be obtained analytically as $\alpha_\mathrm{c} = \frac{2}{c}$. Similar to the polymer Potts model, it can be seen that the CD phase exists as an equilibrium phase when $c$ is large. This is an effect of $c$ since there are only SD and CO phases when $c=0$. 

\end{document}